\documentclass[aps,prl,amsmath,amssymb, twocolumn]{revtex4}

\usepackage{graphicx}
\usepackage{dcolumn}
\usepackage{bm}
\usepackage{textcomp}
\usepackage{amsmath}

\begin{document}

\title{Observation of dressed excitonic states in a single quantum dot}

\author{Gregor Jundt, Lucio Robledo, Alexander H\"ogele, Stefan F\"alt, and Atac Imamo\u{g}lu}
\affiliation{Eidgen\"ossische Technische Hochschule, Institute of Quantum Electronics, Wolfgang-Pauli-Strasse 16, 8093 Z\"urich, Switzerland}

\date{\today}

\begin{abstract}
We report the observation of dressed states of a quantum dot. The optically excited exciton and
biexciton states of the quantum dot are coupled by a strong laser field and the resulting spectral
signatures are measured using differential transmission of a probe field. We demonstrate that the
anisotropic electron-hole exchange interaction induced splitting between the x- and y-polarized
excitonic states can be completely erased by using the AC-Stark effect induced by the coupling field,
without causing any appreciable broadening of the spectral lines. We also show that by varying the
polarization and strength of a resonant coupling field, we can effectively change the polarization-axis
of the quantum dot.
\end{abstract}

\maketitle

Interaction of semiconductor quantum dots (QD) with laser fields has emerged as a new paradigm in
quantum optics. Various experimental achievements based on non-resonant laser excitation of QDs
include the demonstration of single-photon sources \cite{Michler,Moreau}, strong-coupling cavity-QED
\cite{Hennessy,Yamamoto}, and polarization-entangled photon generation \cite{Stevenson1,Gershoni}. In
parallel, resonant excitation of QDs has enabled the first observation of single-QD absorption
\cite{Hogele1} and high-fidelity electron spin-state preparation \cite{Atature}. Signatures of
non-perturbative laser coupling on QD resonances, such as power broadening in absorption line shape \cite{Hogele1}, or Rabi oscillations in photocurrent \cite {Zrenner} have also been
demonstrated. In contrast, the success of pump-probe-type experiments where the effect
of a strong laser field on the QD is studied using resonance fluorescence have been limited, due to
background scattered light arising from the (imperfect) solid-state environment. Two-color
experiments where a weak laser probes the signatures of strong field coupling are in turn limited by
the shot-noise of the strong laser impinging on the same detector as the probe laser. Only a few experiments have successfully overcome these difficulties: resonant fluorescence was detected in
wave-guide geometry \cite{Muller,Melet} which avoids scattered laser background, and transmission of
a weak probe laser was detected using polarization selective rejection of the strong laser \cite{Xu}.
The latter experiment lead to the observation of power-dependent Autler-Townes splitting as well as
signatures of nonlinear Mollow spectrum.

Here, we report an observation of dressed states of a QD, by coherently driving the biexciton
transition with a strong laser field and monitoring the transmission of a weak probe field applied on the fundamental exciton transition. The energy difference of the two lasers is substantial enough to allow for
spectral filtering which in turn provides access to the polarization degrees of freedom. When the
strong field is polarized along one of the QD axes and is tuned off-resonance, its principal effect is
to induce an AC Stark-shift on the co-polarized excitonic state: we demonstrate that this AC
Stark-shift can be used to eliminate the exciton fine-structure splitting (FSS) - an all-optical
alternative to other post-growth manipulation techniques for eliminating FSS
\cite{Kowalik,Stevenson,Gerardot-fs,Seidl} with the goal of deterministic polarization-entangled photon-pair
generation. Alternatively, when the strong laser is not parallel to either of the two QD axes, both
fine-structure split excitonic states can be mixed with the biexciton state in a fully
coherent way: when the Rabi frequency of a resonant strong field exceeds the FSS, the QD axes are
determined by the polarization of the field. The absorption spectrum in this case consists of an
Autler-Townes doublet formed due to coherent mixing of the biexciton with the bright-exciton that is co-polarized with the strong field, and an orthogonally-polarized (uncoupled) dark-exciton.

The energy scheme of a neutral QD and its excited states  is presented in Fig.~1a. The two
near-degenerate neutral exciton states, $|X\rangle$ and $|Y\rangle$, are split by the anisotropic
electron-hole exchange energy which gives rise to the FSS $\hbar \delta_{xy}$ \cite{Gammon,Zunger}.
The exciton states couple via linearly polarized optical transitions to the QD ground state
$|0\rangle$ (empty QD) and the biexciton state $|XX\rangle$ (doubly excited QD). The corresponding
transition energies differ by $\sim 3.5$~meV due to Coulomb-interaction of the electrons and holes
confined in the QD: this anharmonicity allows us to probe and manipulate the exciton and biexciton
states independently and profit from spectral filtering through the combination of a reflective
holographic grating, an imaging lens and a $100~\mu$m diameter pinhole. We choose the energy
$(\mathrm{E}_{x}+\mathrm{E}_{y})/2$ of a QD without FSS as the reference for energy detunings.

Optical control is established with two narrow-band diode  lasers with adjustable frequencies and
polarization vectors, $\hat{u}_c$ and $\hat{u}_p$. The corresponding laser intensities
determine the Rabi frequencies, $\Omega_c$ and $\Omega_p$, through the electric field $E \hat{u}_i$
along the exciton dipole moment $\mu \hat{u}_{j}$ ($ j= x, y$) through $\Omega_i=(\hat{u}_{j} \cdot
\hat{u}_i)\mu E /\hbar$ ($i = c,p$). The charging state of the QD and the electrostatic DC
Stark-field are defined using a field-effect heterostructure where the QDs are sandwiched between a
semitransparent Schottky gate electrode and a highly $n$-doped Ohmic back contact \cite{Warburton1}.
The sample was operated in an optical helium bath cryostat at $4.2$~K base temperature and excited
through a diffraction limited spot with $1.5~\mu$m full-width at half-maximum.

We identify the neutral exciton state of a single QD either from gate-controlled charging plateaus in
photoluminescence \cite{Warburton1} or differential absorption spectra which show characteristic
fine-structure split resonances \cite{Hogele1}. Differential absorption was measured using DC
Stark-shift modulation spectroscopy \cite{Alen} with a fixed probe laser frequency $\Omega_p$.
Fig.~1b shows the corresponding spectrum (black solid line) at 4.2~K with the probe field applied at
the exciton transition: both $|X\rangle$ and $|Y\rangle$ exciton resonances separated by the FSS
$\hbar \delta_{xy}=20~\mu$eV are detected since the probe laser polarization was chosen to have equal
overlap with the polarization axes of the QD, $\hat{\pi}_x$ and $\hat{\pi}_y$ ($\hat{u}_p =
\hat{\pi}_{+45}$).

\begin{figure}
\includegraphics[scale=1.0]{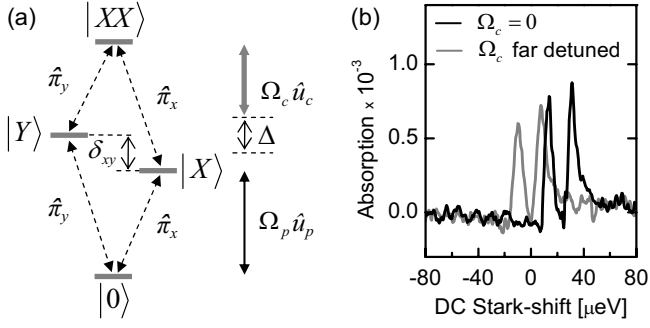}
\caption{\label{fig:energy-scheme} (a) Energy scheme of the quantum dot optical transitions coupled
by polarized laser fields of Rabi frequency $\Omega_{P}$ (weak probe laser) and $\Omega_{c}$ (strong
coupling laser) with the corresponding polarization vectors $\hat{u}_p $ and $\hat{u}_c$. The exciton
states $|X\rangle$ and $|Y\rangle$, split by fine-structure $\delta_{xy}$ due to anisotropic exchange,
couple through linearly polarized transitions $\hat{\pi}_x$ and $\hat{\pi}_y$ to the biexciton state
$|XX\rangle$ and the empty quantum dot ground state $|0\rangle$. The detuning $\Delta$ of the
coupling laser is measured with respect to the energy $\mathrm{E}=1/2\cdot(\mathrm{E}_x+\mathrm{E}_y)$ of a neutral exciton without fine-structure splitting. (b) Differential transmission spectra of the neutral exciton transitions in a single quantum dot probed with circularly polarized laser at $4.2$~K: The two
spectra were measured for the same dot with the coupling laser on but far detuned (gray solid line)
and in the absence of a coupling laser (black solid line).}
\end{figure}

When we turn the coupling laser on, we observe a blue-shift of the exciton energies by $20$ to $30~\mu$eV
(Fig.~1b red curve). This shift is independent of the coupling laser frequency, provided that the
detuning of the coupling laser from the QD resonances is much larger than the width of the lines and the
Rabi frequencies. Similar order-of-magnitude blue-shifts in absorption lines, most likely induced by
creation and trapping of carriers in deep defects within the QD environment, have been observed in
other experiments based on strong laser excitation of QDs \cite{Xu}. For the QD that we studied in
our experiments, we measured a QD exciton transition linewidth of $\gamma_0=5~\mu$eV for probe
laser intensities well below saturation. This linewidth, which remained unchanged when the
off-resonance coupling field was turned on, is not lifetime limited ($\gamma_{rad}=1~\mu$eV) and is
most likely to be induced by the spectral fluctuations of the exciton energy \cite{Hogele1}. All of
the experiments reported here were carried out in the limit where the excitonic resonances were
weakly power-broadened to $\gamma \simeq 9~\mu$eV by the probe laser with $\Omega_p\simeq 1.1\,
\gamma_0$, in order to maximize the signal-to-noise ratio of differential transmission measurements
\cite{Gerardot}.

\begin{figure}
\includegraphics[scale=1.0]{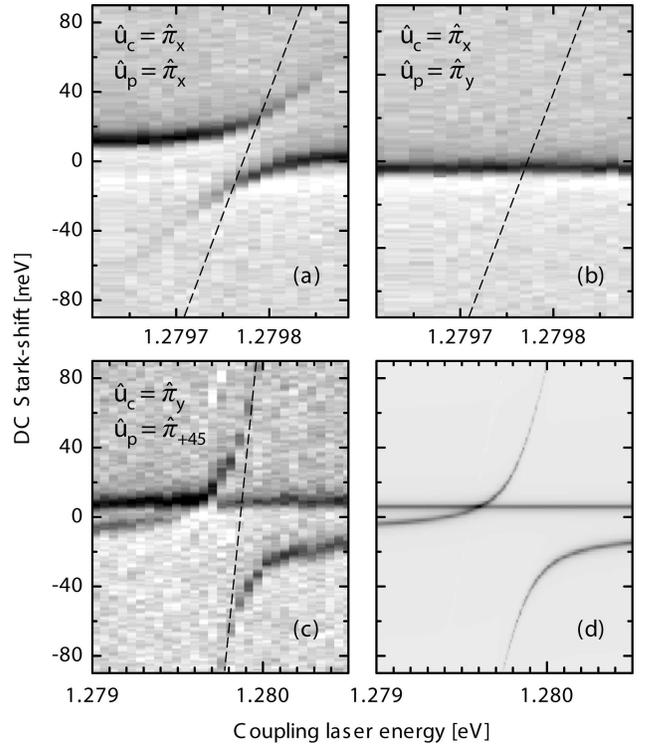}
\caption{\label{fig:XY} Neutral exciton states in the presence of a strong exciton-biexciton coupling field in grey scale representation (black and white corresponding to an absorption of $0.7\times10^{-3}$ and 0, respectively). Co-polarized probe and coupling in (a) reveal the Autler-Townes doublet and the AC Stark-shift of the neutral exciton $|X\rangle$ that is parallel to the coupling laser polarization. The exciton state $|Y\rangle$ orthogonal to the coupling laser remains unaffected, as shown for cross-polarized probe and coupling in (b). Experimental data (c) showing that the exchange splitting can be canceled by exclusively shifting one transition (note  different abscissa scales in the upper and lower panel). The corresponding calculation is plotted in (d). The DC Stark-shift affects both the exciton and the biexciton energies equivalently and the dashed lines of slope 1 indicate the resonance condition for the exciton-biexciton coupling laser. Parameters: $\Omega_c=45~\mu$eV in Fig.s 2 (a), (b) and $\Omega_c=120~\mu$eV in Fig.s 2(c) and (d); $\Omega_p=5.3~\mu$eV in all graphs.}
\end{figure}

Figure 2a,b presents experimental data where a coupling field with $\Omega_{c} = \Omega_{c,x} = 45~\mu$eV
polarized along one of the QD axes ($\hat{\pi}_x$) is tuned across the biexciton-exciton transition:
the grey-scale plots depict experimental absorption spectra measured as a function of the coupling laser
frequency (horizontal axis) and the DC Stark-shift controlled detuning with respect to the (fixed frequency) probe laser (vertical axis). Since the exciton and the biexciton exhibit different DC Stark-shifts, the detuning of the coupling field from the biexciton transition varies as we change the gate voltage. The dashed lines in Fig.~2 indicate the resulting dependence of the exact biexciton-coupling laser resonance condition on the DC Stark shift of the exciton line. For a probe laser that is polarized parallel to the coupling laser, we observe an Autler-Townes doublet, typical for ladder-type atomic systems (Fig.~2a). In contrast, an orthogonally polarized probe laser coupling to the $\hat{\pi}_y$-polarized exciton transition shows no dependence on the coupling laser detuning (Fig.~2b): these results demonstrate that it is possible to realize highly selective coherent manipulation of QD resonances.

Figure 2c shows the absorption spectrum of a probe laser that is $\hat{\pi}_{+45}$-polarized and hence
couples to both excitonic lines simultaneously. For a $\hat{\pi}_x$-polarized coupling field with a
Rabi frequency $\Omega_{c,x} = 120 \mu$eV, we find that the dressed $|X\rangle$-polarized QD state
$|D_U\rangle$ becomes degenerate with the $|Y\rangle$-exciton for a detuning of $260 \mu$eV. The
vanishing absorption of the other (lower energy) dressed state indicates that $|D_U\rangle$ is
predominantly exciton-like, with vanishing probability for biexciton excitation. In this limit, the
effect of the coupling laser can be understood as creating an AC-Stark-shift of the $|X\rangle$ state
that exactly cancels the FSS. The experiment depicted in Fig.~2c agrees very well with calculations
shown in Fig.~\ref{fig:XY}d, which are in turn based on the Hamiltonian (expressed in the basis
$|X\rangle, |Y\rangle, |XX\rangle$):
\begin{equation}
H = -\frac{\hbar}{2} \cdot
\begin{pmatrix}
\delta_{xy}  &     0             &    \Omega_{c,x} \\
0            &    -\delta_{xy}   &    \Omega_{c,y} \\
\Omega_{c,x} & \Omega_{c,y}      & -2 \cdot \Delta
\end{pmatrix}
\label{eq1}
\end{equation}
Here, $\hbar \Delta$ is the detuning between the coupling laser energy and the exciton-biexciton
transition, and $\Omega_{c,x}$, $\Omega_{c,y}$ are the Rabi frequencies of the coupling laser along
the exciton dipole moment axes $\hat{\pi}_x$ and $\hat{\pi}_y$. The calculations depicted in Fig.~2d
use Rabi frequencies derived from the experimentally measured coupling laser intensity and previously measured values of the excitonic oscillator strength ($f=10$)\cite{Zrenner}. 
The Hamiltonian of Eq.~(\ref{eq1}), derived using the electric-dipole and rotating-wave approximations,
accurately describes the dressed excited states of the QD in the limit of a perturbative probe field.
Since the experimental results are obtained using a probe field with $\Omega_p\simeq 1.1\, \gamma_0$,
we could only expect to have a qualitative match with the theoretical model. The effect of
dissipation and line broadening can be included using a master equation in the Lindblad form.

\begin{figure}
\includegraphics[scale=1.0]{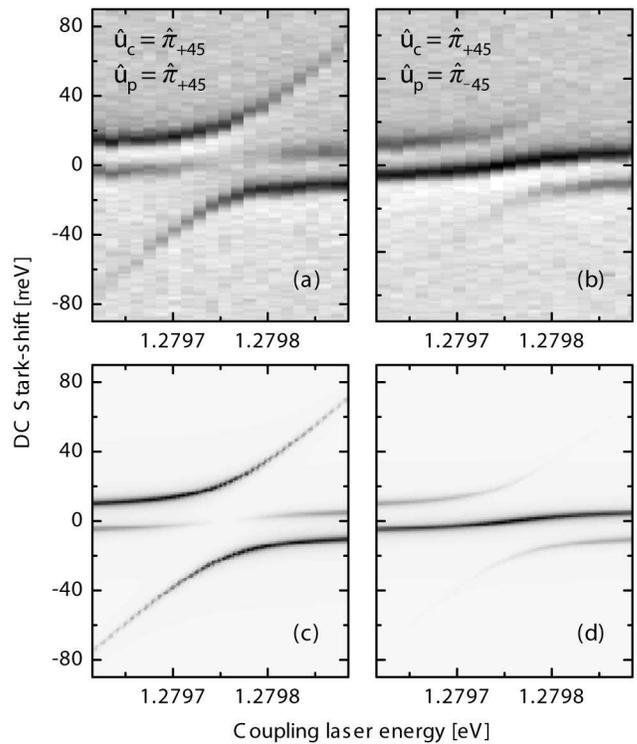}
\caption{\label{fig:coupling}The strong laser is coupled equally  to the two transitions. The upper panel shows experimental data for (a) co-polarized lasers and (b) cross-polarized lasers. The lower panel shows the corresponding simulation with $\Omega_{c,x}=\Omega_{c,y}=40~\mu$eV and $\Omega_p=5.3~\mu$eV which follow from experimental settings. The grey scales are the same as in Fig.~2.}
\end{figure}

When the coupling laser frequency is not polarized along one of the QD axes, it leads to the
formation of three dressed-states which are coherent superpositions of the biexciton state and the
two exchange-split excitonic states; in this regime, laser polarization and the competition between
the anisotropic exchange interaction and laser Rabi coupling determine the optical response (see
Eq.~(\ref{eq1}). In the limiting case of $\Omega_{c,x}, \Omega_{c,y} \gg \delta_{xy}, \Delta$, we
would expect the exchange interaction to be insignificant. Among the three dressed-eigenstates, one
will be a superposition of the two bare excitonic states that lead to absorption/emission polarized
orthogonal to the coupling laser; this dark-eigenstate \cite{footnote} will have an energy that is equal
to the bare exciton emission energy (i.e. the energy when $\delta_{xy} = 0 = \Omega_{c}$). The other
two eigenstates will be superpositions of the exciton and biexciton states that are split by
$\Omega_{c}$, much like in the case depicted in Fig.~\ref{fig:XY}.

Figure~\ref{fig:coupling} shows an experiment where the coupling laser is polarized at $45 ^{\circ}$
with respect to the QD axis. By applying a coupling laser of $1.05 ~kW/cm^{2}$ intensity, we achieve $\Omega_{c,x} = \Omega_{c,y} = 40 \mu$eV $> \hbar \delta_{xy}$. Figure ~\ref{fig:coupling} shows the evolution of the QD
excitonic absorption as the coupling field frequency is swept through the biexciton-exciton resonance.
Unlike Fig.~\ref{fig:XY}, we observe that all three states are coupled; in particular, the state that
is X-polarized for large red-detuning maps on to the Y-polarized exciton state for large
blue-detuning. For $\Delta = 0$, this eigenstate can be written as $|X\rangle - |Y\rangle$, i.e. its polarization axes is orthogonal to that of the coupling field. In this regime, the QD response can be
characterized as consisting of an Autler-Townes doublet polarized parallel to the coupling laser and
an orthogonally polarized dark-excitonic state.

An important question that arises for emitters embedded in a solid-state matrix is whether strong
laser excitation leads to unwanted coupling or broadening effects that cannot be captured by the
simple 4-level model depicted in Fig.~1a. To address this question, we have measured the total area
under the absorption resonances and the transition linewidths, as a function of the coupling laser
intensity. While the former tells us whether there is any appreciable excitation of other QD states,
the latter would point out to decoherence effects mediated by the strong field itself. We find that
even for the strongest coupling field intensities used in our experiments ($\Omega_{c,x} = 120~\mu$eV), the changes in the total area under the resonances were well within the error-bar of our
experiments. We did observe a slight increase of $\sim 10\%$ in the average transition linewidths
when the coupling laser was on resonance. More importantly, we observed that the fluctuations in the
measured linewidths (from one scan to another) was enhanced; for example, the measured
linewidths of the two Autler-Townes split lines could differ by as much as $\sim 50\%$.
We tentatively conclude that these fluctuations as well as the slight linewidth enhancement is due to
intensity fluctuations of the coupling laser, and is not a result of a deviation from the simple
4-level model.

In summary, we have demonstrated that QD states can be coherently manipulated, by applying a
non-perturbative resonant laser field on the biexciton-exciton transition. The observation of polarization-selective AC-Stark shift of excitonic resonances could be considered as
a key step towards laser-induced Zeeman-like shift of QD spin states, which would in turn allow for
fast manipulation of spin degrees-of-freedom \cite{Imamoglu}. The excitonic dark-states that are immune to strong
field excitation suggest that phenomena such as coherent population trapping are within reach in
these solid-state emitters.

\begin{acknowledgments}
G. J. acknowledges Yong Zhao for many fruitful discussions. This work is supported by NCCR Quantum Photonics (NCCR QP), research instrument of the Swiss National Science Foundation (SNSF), and EU RTN EMALI.
\end{acknowledgments}


\begin{thebibliography}{}
\bibliographystyle{unsrt}

\bibitem{Michler} P. Michler \textit{et al.}, Science {\bf 290}, 2282 (2000).

\bibitem{Moreau} E. Moreau \textit{et al.}, Appl. Phys. Lett. {\bf 79}, 2865 (2001).

\bibitem{Hennessy} K. Hennessy \textit{et al.}, Nature (London) {\bf 445}, 896 (2007).

\bibitem{Yamamoto} D. Press \textit{et al.}, Phys. Rev. Lett. {\bf 98}, 117402 (2007).

\bibitem{Stevenson1}  R. M. Stevenson \textit{et al.}, Nature (London) {\bf 439}, 179 (2006).

\bibitem{Gershoni} N. Akopian \textit{et al.}, Phys. Rev. Lett. {\bf 96}, 130501 (2006).

\bibitem{Hogele1} A. H\"{o}gele \textit{et al.}, Phys. Rev. Lett. {\bf 93}, 217401 (2004).

\bibitem{Zrenner} A. Zrenner \textit{et al.}, Nature (London) {\bf 418}, 612 (2002).

\bibitem{Atature} M. Atat\"{u}re  \textit{et al.}, Science {\bf 312}, 551 (2006).

\bibitem{Xu}  X. Xu \textit{et al.}, Science {\bf 317}, 929 (2007).

\bibitem{Muller}  A. Muller \textit{et al.}, Phys. Rev. Lett. {\bf 99}, 187402 (2007).

\bibitem{Melet}  A. Melet \textit{et al.}, arXiv:0707.3061v1.

\bibitem{Kowalik} K. Kowalik \textit{et al.}, Appl. Phys. Lett. {\bf 86}, 041907 (2005).

\bibitem{Gerardot-fs} B. Gerardot \textit{et al.}, Appl. Phys. Lett. {\bf 90}, 041101 (2007).

\bibitem{Stevenson} R. M. Stevenson \textit{et al.}, Phys. Rev. B {\bf 73}, 033306 (2006).

\bibitem{Seidl} S. Seidl \textit{et al.}, Appl. Phys. Lett. {\bf 88}, 203113 (2006).

\bibitem{Gammon} D. Gammon \textit{et al.}, Phys. Rev. Lett. {\bf 76}, 3005 (1996).

\bibitem{Zunger} G. Bester, S. Nair, and A. Zunger, Phys. Rev. B {\bf 67}, 161306(R) (2003).

\bibitem{Warburton1}  R. J. Warburton \textit{et al.}, Nature (London) {\bf 405}, 926 (2000).

\bibitem{Alen} B. Al\'{e}n \textit{et al.}, Appl. Phys. Lett. {\bf 83}, 2235 (2003).

\bibitem{Gerardot} B. Gerardot \textit{et al.}, Appl. Phys. Lett. {\bf 90}, 221106 (2007).

\bibitem{Imamoglu} A. Imamoglu \textit{et al.}, Phys. Rev. Lett. {\bf 91}, 017402 (2003).

\bibitem{footnote} We emphasize that this excitonic state is dark in the sense of not having any biexciton component.

\end{thebibliography}
\end{document}